\documentclass[11pt]{article}
\usepackage{moriond,epsfig}

\def\be{\begin{equation}}
\def\ee{\end{equation}}
\def\bea{\begin{eqnarray}}
\def\eea{\end{eqnarray}}

\begin{document}
\vspace*{4cm}
\title{MULTIWAVELENGTH OBSERVATION OF SS\,433 JET INTERACTION WITH THE INTERSTELLAR MEDIUM}

\author{ Y. FUCHS }

\address{Service d'Astrophysique, CEA/Saclay, Orme des Merisiers b\^at. 709\\
91191 Gif-sur-Yvette, France}

\maketitle\abstracts{
SS\,433 is an X-ray binary emitting persistent relativistic double sided
jets that expand into the surrounding W50 radio nebula. The SS\,433\,/\,W50
system is then an excellent laboratory for studying relativistic jet
interaction with the surrounding interstellar medium. In this context,
part of W50 nebula has been mapped with ISOCAM at 15 micron, where the
large scale X-ray jets are observed. I will show the results,
particularly on the W50 western lobe, on two emitting regions detected
in IR with IRAS, and observed in millimeter wavelength (CO(1-0)
transition).
It is uncertain whether these regions are
due to heated dust by either young stars or the relativistic jet, or
to synchrotron emission from shock re-acceleration regions as in some
extragalactic jets from AGN. In this latter case, INTEGRAL might
detect soft gamma-ray emission.}

\section{Introduction: the SS\,433\,/\,W50 system}

	\indent SS\,433 is an X-ray binary composed of probably a
	massive star and probably a neutron star, but the nature of
	none of them has been confirmed. The binary system emits
	compact relativistic jets observed at subarcsecond scales in
	radio, and which characteristic is to show a precession
	movement. The Doppler shifted lines observed in the optical
	spectrum enable to calculate the precession model parameters,  
	resulting in relativistic ($v=0.26$\,c) jets covering a
	cone with a half opening angle of $\theta=19.8^\circ$, which
	axis has an inclinaison angle of $i=78.82^\circ$ to the line
	of sight, and a precession period of about 162.5 days, the
	binary period being close to 13.08 days (Margon\,\cite{Margon}
	1984). The feature of these precessing jets is that during a
	small time interval of the precession, the eastern jet which
	is approaching most of the time recedes, and the western jet
	approaches instead of receding.\\

\begin{figure}[!ht]
\centerline{\psfig{figure=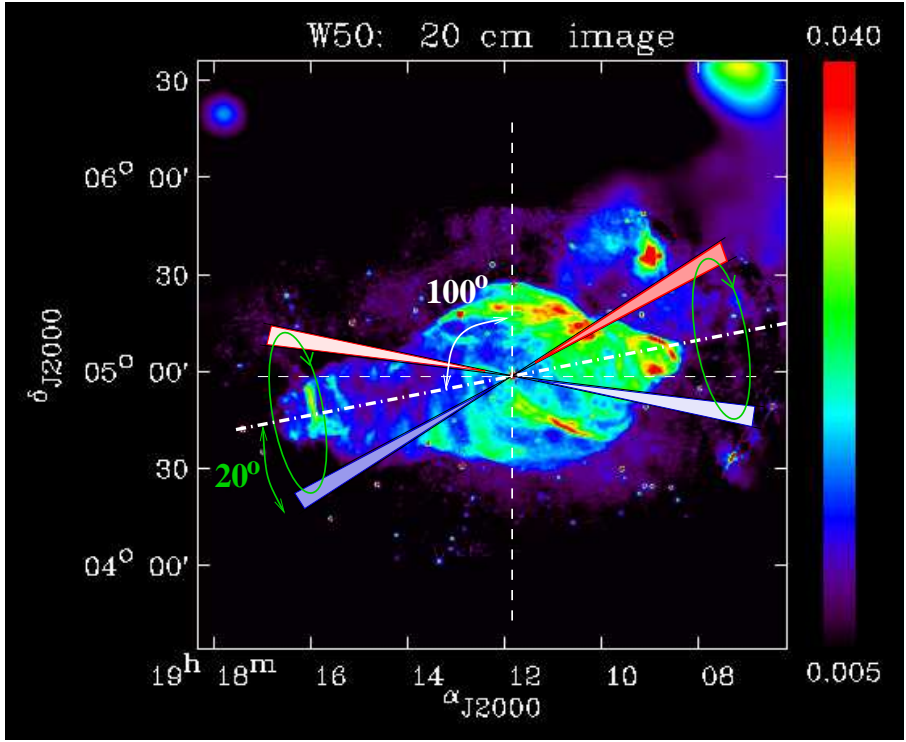,width=12cm}}
\caption{W50 at 20\,cm with the projected precession cone on the sky plane. The cone axis has a position angle of $100^\circ$ according to Hjellming and Johnston (1981). The two radio lobes lie within this cone.}
\label{figW50cone}
\end{figure}

	   \indent This binary is the center of the ``sea-shell'' radio
	nebula W50 shown at 20\,cm (from Dubner {\em et al.}\cite{Dubner}
	1998) in the figure~\ref{figW50cone}. This $2^\circ \times
	1^\circ$ nebula ($\sim 120$\,pc $\times$ 60\,pc at a distance
	of 3.5\,kpc) has a circular central shape considered as a
	supernova remnant, with two extensions called lobes or
	``wings'' resulting in this sea-shell or goose beak form. The
	eastern wing exhibits a clear helical pattern which mirrors at
	large scales the precession of the jets from SS\,433. The
	western wing, smaller and brighter, appears to interact with a
	denser medium. So different ambient conditions may result in
	different acceleration and emission mechanisms occurring inside
	the remnant. On figure~\ref{figW50cone}, the projection of the
	precessing movement on the sky plane is superimposed on the
	20\,cm image of the nebula showing that these wings are well
	constrained inside the precession cone. Thus W50 structure
	reveals the connection between the subarcsecond relativistic
	jets from SS\,433 and the extended nebula over $\sim 5$ orders
	of magnitude in scale.

\section{Multiwavelength observations}
\subsection{ISOCAM}
	\indent In order to prove that the W50 elongated shape is due
	to jets, we observed it with ISOCAM, the infrared camera on
	board of the Infrared Space Observatory (ISO), at 15\,$\mu$m
	(large band filter 14-16\,$\mu$m) with a spatial resolution of
	$6''\times6''$per pixel. We could not map the whole nebula so
	we mapped a small part of the eastern lobe where an X-ray knot
	lies, the north-east quarter of the central circular part of
	W50, and nearly all the western lobe.

  \indent There is no particular 15\,$\mu$m emission in the observed eastern
   parts of W50, and no correlation was found between the IR images
   and the corresponding ones in X-ray (with ROSAT and ASCA) and radio (at
   20\,cm), both for the central field and for the X-ray knot
   area. This is not surprising as the central part of W50 must have
   been swept away from its material by the supernova explosion, and
   the eastern lobe is described as faint in radio and less dense than
   the western one by Dubner {\em et al.}\cite{Dubner} (1998). The X-ray knot, the
   brightest knot seen by ROSAT at 0.1--2.4\,keV and called ``e2'' by
   Safi-Harb and \"Ogelman\,\cite{Safi97} (1997), is coincident with an optical
   filament and is probably due to shocks with the supernova remnant
   (Safi-Harb and Petre\,\cite{Safi99} 1999), but it is not visible at 15\,$\mu$m
   with ISOCAM sensitivity.

\begin{figure}[!ht]
\centerline{\epsfig{figure=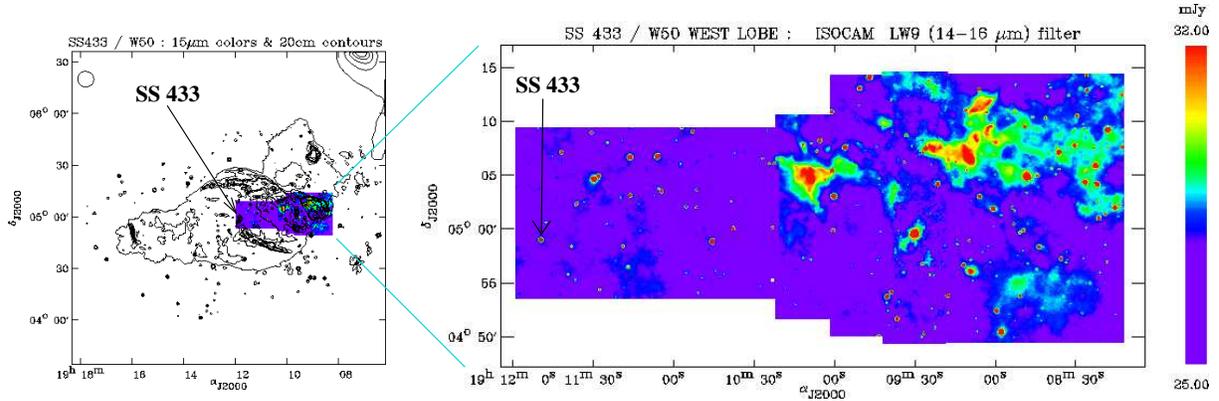,width=16cm}}
\vspace*{-0.2cm}\caption{Left: location of the ISOCAM map in the W50 western lobe; W50 is shown 
at 20\,cm in contours. Rigth: ISOCAM 14--16\,$\mu$m image with a $-\sigma$ to +2\,$\sigma$ scale around the median flux value.
\label{figobsW50}}
\end{figure}

	\indent Thus in the following we will concentrate on the
	western lobe. Its map at 14--16\,$\mu$m shown on
	figure~\ref{figobsW50} reveals infrared structures at
	distances $>20'$ away from SS\,433, consisting in faint
	extended emissions and ``hotspots''.  The two main emitting
	regions are aligned with the direction of the precession cone
	axis. The IR emission zones may indicate denser region inside
	the western lobe partly hit by the relativistic jet. In order
	to understand these IR emissions, let us draw a comparison with
	other wavelength observations.

\vspace*{-0.6cm}\subsection{radio}
\begin{figure}[!ht]
\centerline{\psfig{figure=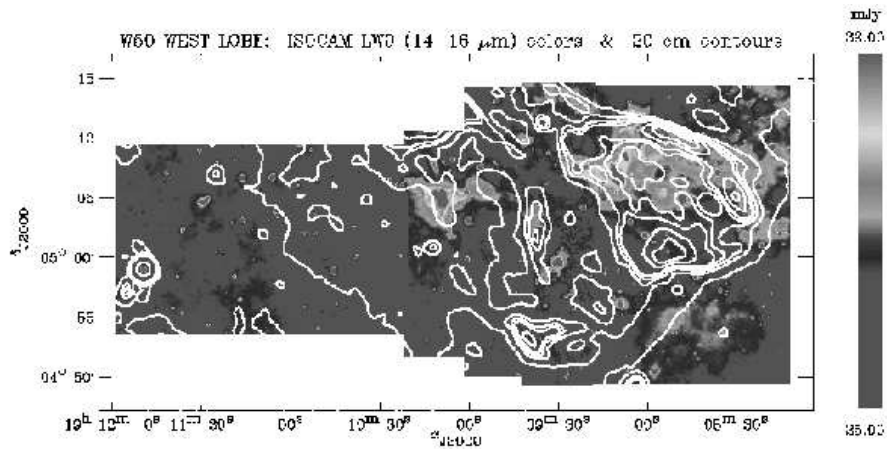,width=12cm}}
\vspace*{-0.2cm}\caption{Comparison of ISOCAM 15\,$\mu$m image of W50 western lobe with the 20\,cm contours.
\label{figIRradio}}
\end{figure}

	\indent Figure~\ref{figIRradio} shows the superimposition of the 20\,cm
	contours on the ISOCAM image. The radio and 15\,$\mu$m
	emissions are partly spatially coincident, only at the edge of
	the western ``ear'' of W50, where the radio emission is the
	strongest. There may be no IR emission souther because of a
	less dense medium.
\vspace*{-0.3cm}

\vspace*{-0.5cm}\subsection{X-ray}
\begin{figure}[!ht]
\centerline{\psfig{figure=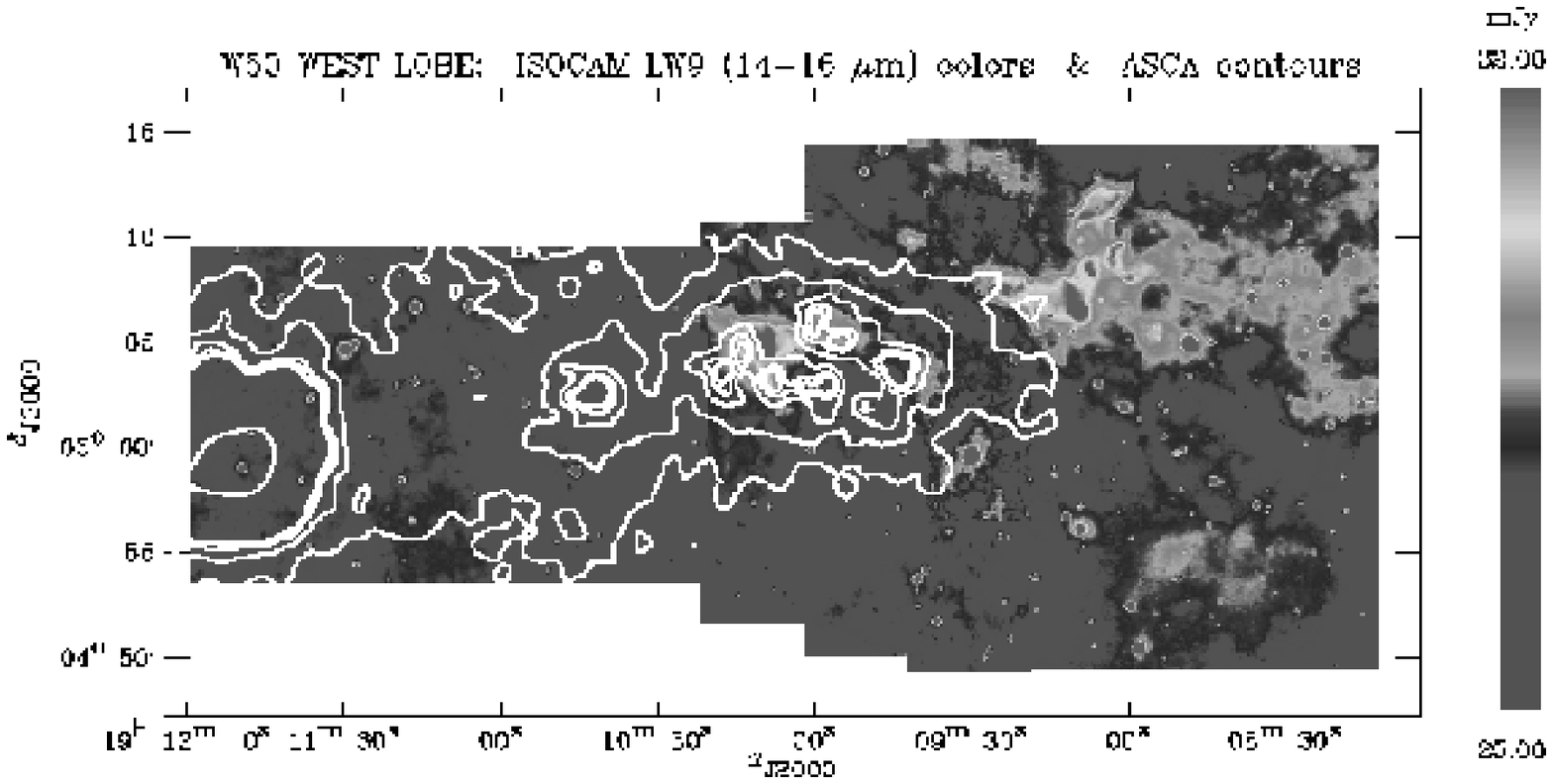,width=12cm}}
\centerline{\psfig{figure=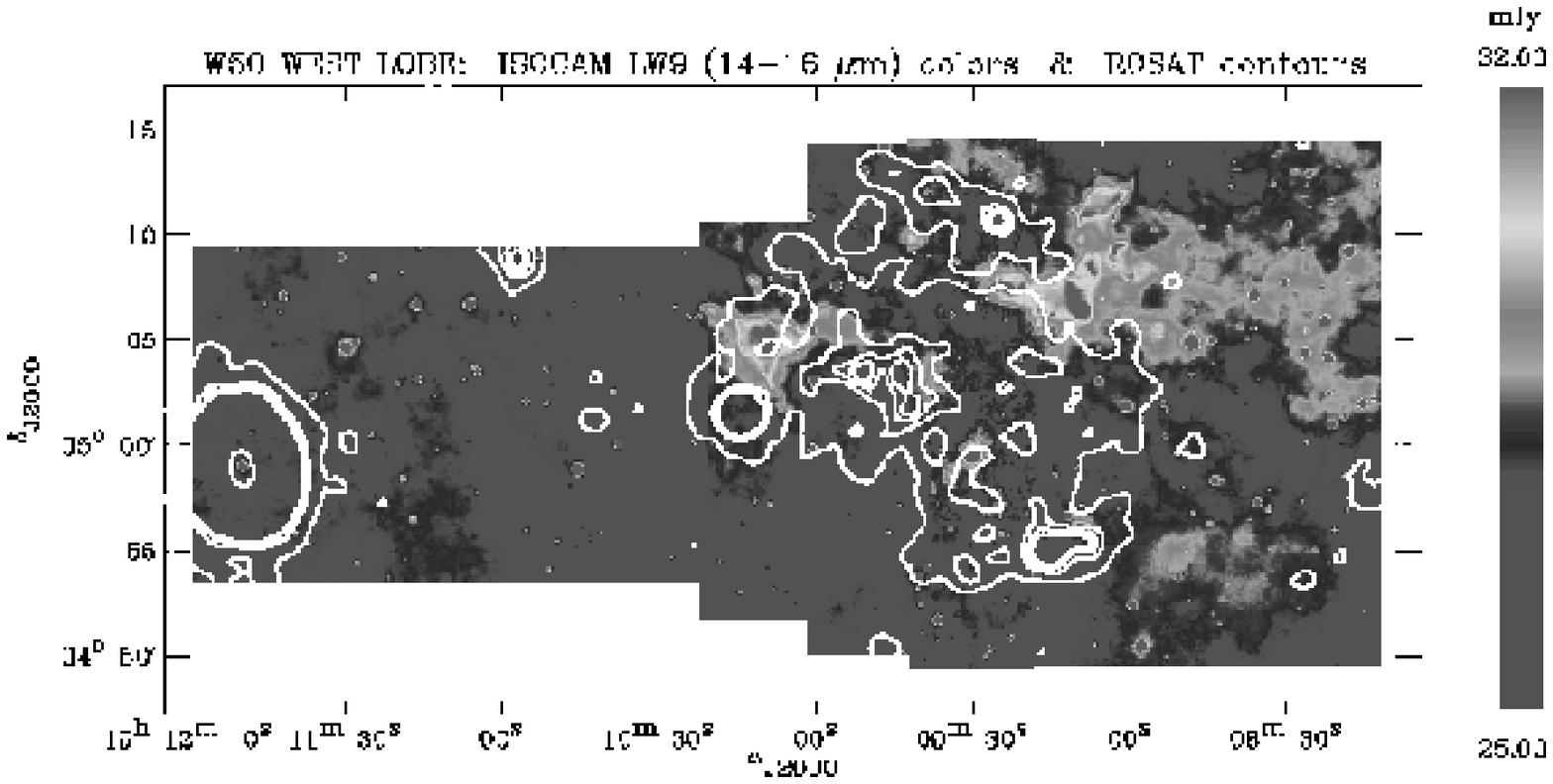,width=12cm}}
\vspace*{-0.2cm}\caption{Comparison between ISOCAM 15\,$\mu$m image of W50 western lobe and ASCA 0.5-9\,keV (top) and ROSAT 0.11-2.35\,keV (bottom) contour images. Spatial resolution are $6'' \times 6''$ per pixel for ISOCAM, $15'' \times 15''$ for ASCA and $12'' \times 12''$ for ROSAT, these latter being smoothed with a $1'$ FWHM Gaussian.
\label{figIRX}}
\end{figure}

	\indent The western X-ray lobe was observed with ROSAT
	(0.1--2.4\,keV) by Brinkmann {\em et al.}\cite{Brink} (1996),
	and with ROSAT and ASCA (0.5--9\,keV) by Safi-Harb and
	\"Ogelman\,\cite{Safi97} (1997) who could not conclude if its
	emission is thermal or not. X-ray lobes are found to fill the
	gap between SS\,433 and the radio ears, with their spectrum
	generally softening away from SS\,433 along the precession
	jets axis (Safi-Harb and Petre\,\cite{Safi99} 1999). The hard
	X-ray emission detected with ASCA (0.5-9\,keV) is knotty,
	appearing at $>15'$ from SS\,433, and focused on the
	precession axis cone as the ISOCAM emission (see
	figure~\ref{figIRX} top). Due to ASCA limited field of view
	the edge of the lobe was not observed, but there seems to be a
	spatial anticorrelation between hard X-ray and and the farest
	mid-IR emissions from the binary system. The brightest X-ray
	knot is $31'$ west from SS\,433, where there is 15\,$\mu$m
	emission, but X-ray observation with a better spatial
	resolution is needed to be conclusive.  The soft X-ray
	emission seen by ROSAT (0.11-2.35\,keV) is also knotty and
	diffuse (see figure~\ref{figIRX} bottom). It is faint and
	spatially anticorrelated with the 15\,$\mu$m emission,
	although there is very faint (too faint to be drawn as
	contours on figure~\ref{figIRX}) and very soft X-ray emission
	at the edge of the radio mid-IR emitting ear (Brinkmann {\em et al.}\cite{Brink} 1996).\\ 
	\indent The
	nature of the X-ray emission is still uncertain between
	thermal and non thermal model, the absence of observed thermal
	emission line suggesting that either the plasma is very hot (a
	few keV) or the spectrum is non thermal. It is possible that
	both thermal and non thermal emission mechanisms are present,
	as would be expected from the interaction of a jet with a
	inhomogeneous medium.
	Thus the question is: are the
	X-ray knots particles acceleration \mbox{zones ?} If yes, they should
	be detectable with INTEGRAL.
\vspace*{-0.4cm}

\vspace*{-0.5cm}\subsection{IR and millimetric} \label{sectIRmm}
\begin{figure}[!ht]
\centerline{\psfig{figure=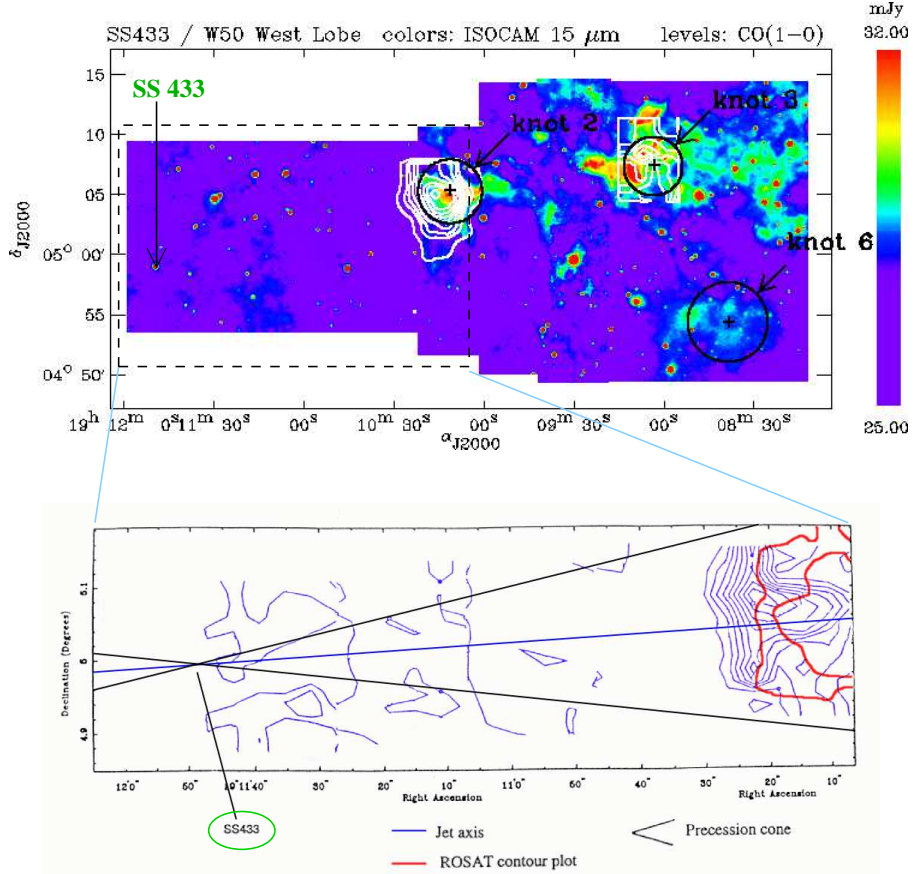,width=12cm}}
\vspace*{-0.2cm}\caption{CO(1-0) emission contours obtained with MOPRA (bottom and knot 3) and SEST (knot 2) telescopes compared to ISOCAM 15\,$\mu$m image of W50 west lobe.
\label{figIRCO}}
\end{figure}

\begin{figure}[!ht]
\centerline{\psfig{figure=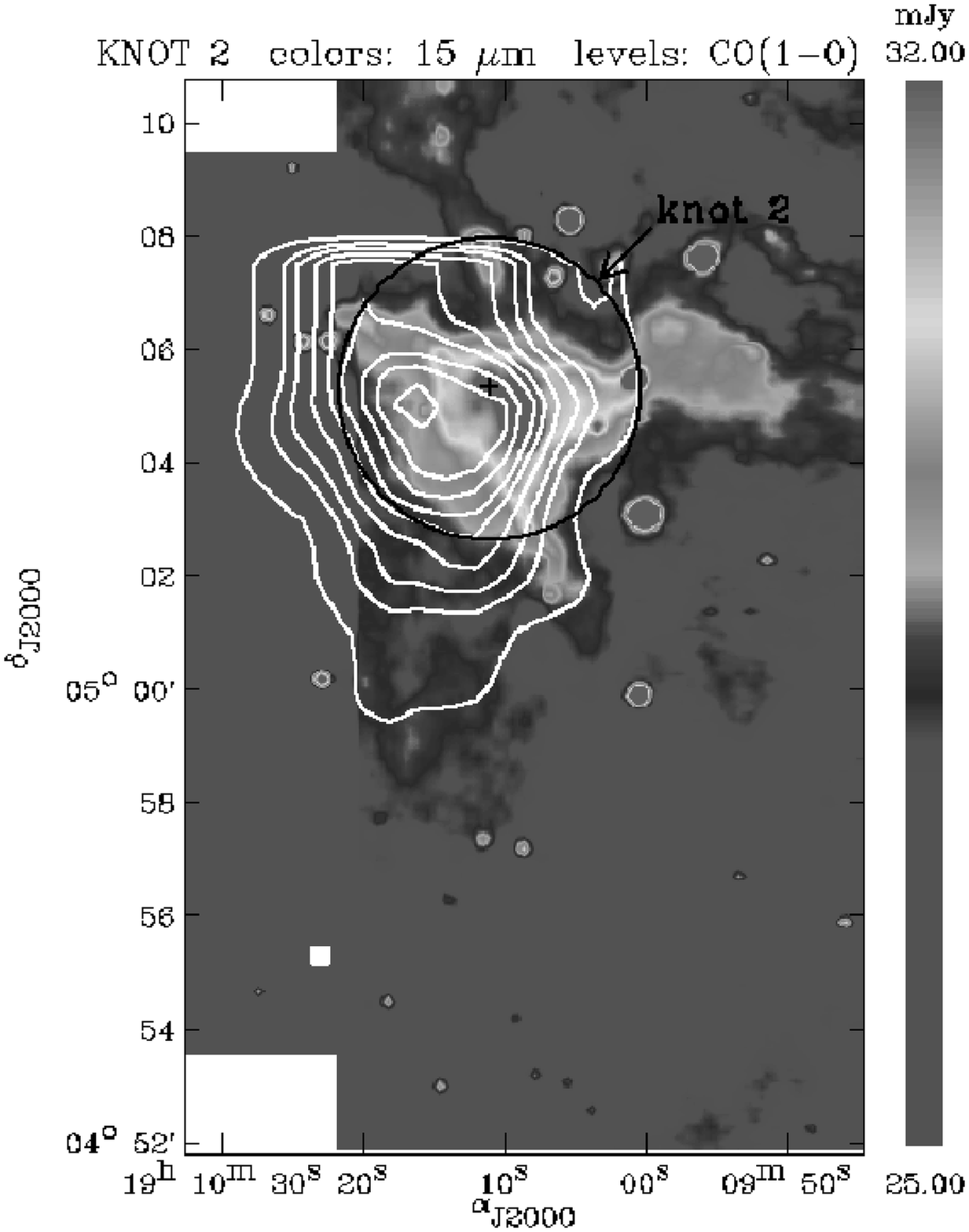,width=5.5cm}\hspace*{0.5cm}
\psfig{figure=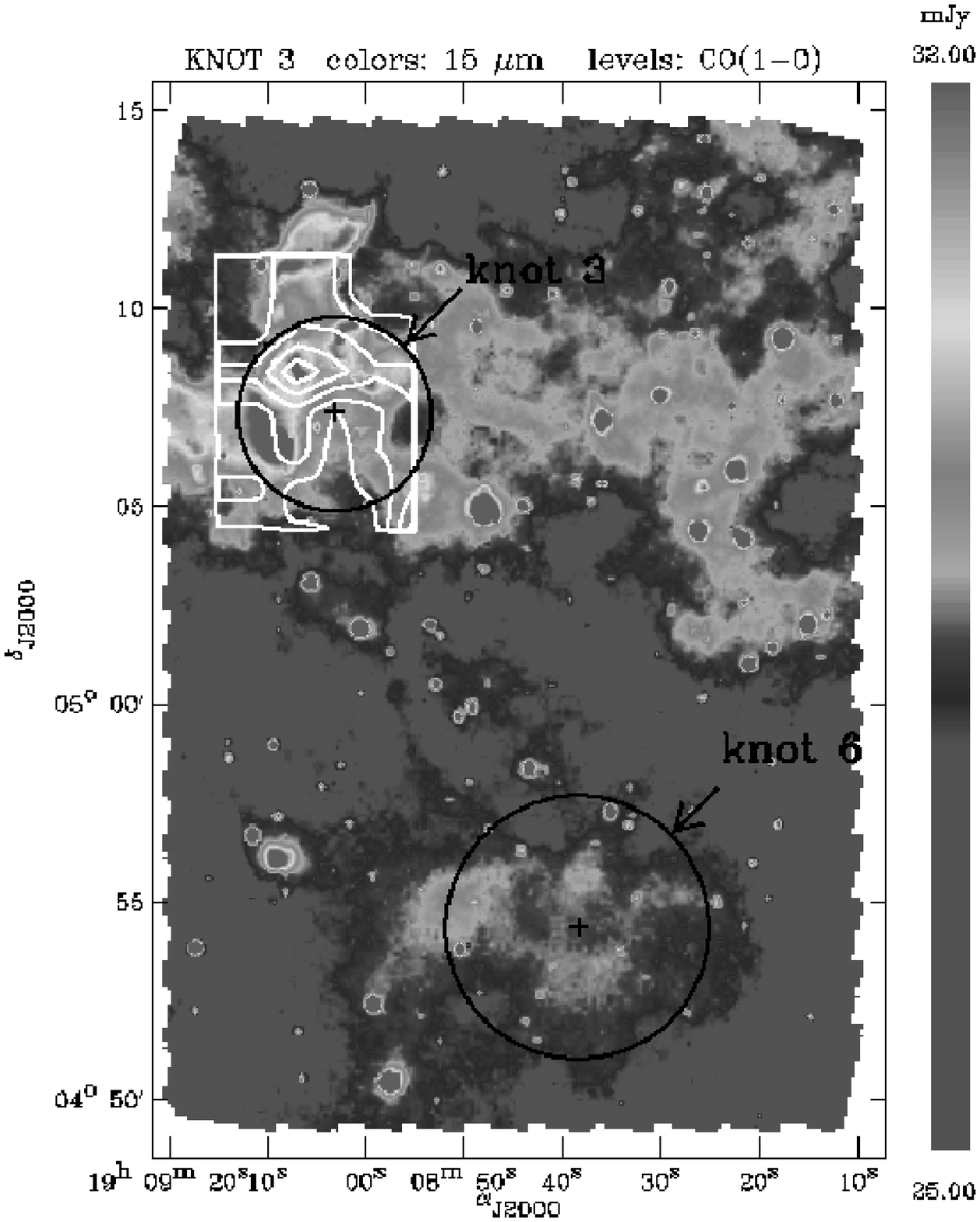,width=6cm}}
\vspace*{-0.2cm}\caption{Zoom on knot 2 and knot 3 regions of figure~\ref{figIRCO}
\label{figIRCOzoom}}
\end{figure}

	\indent W50 had already been observed in IR with the IRAS
	satellite before ISO. Band\,\cite{Band} (1987) thus discovered
	several IR knots at 12, 25, 60 and 100\,$\mu$m, three of them
	being in the ISOCAM field of view as shown on
	figure~\ref{figIRCO} where the circles represents the size of
	these knots as seen by IRAS, and the names are those given by
	Band\,\cite{Band} (1987).  The $6'' \times 6''$ per pixel
	ISOCAM spatial resolution enables to reveal their structure
	for the first time. The 15\,$\mu$m knots emission is diffuse
	with punctual or not ``hot spots''. As ``knot 6'' projection
	onto the sky plane is at the external limit of the W50 radio
	nebula and has probably no link with it, we will concentrate
	on ``knot 2'', the nearest region from SS\,433 and ``knot 3'',
	the farest.\\ 
	\indent W50 western lobe has been observed with the millimetre
	telescope SEST and MOPRA.  No emission from the CO(1-0)
	transition line at 2.6\,mm (115\,GHz) is observed between
	SS\,433 and ``knot 2'', as in IR at 15\,$\mu$m in the ISOCAM
	image (see bottom of figure 5). Two CO(1-0) emitting regions
	are coincident with the two IRAS knots as shown in
	figure~\ref{figIRCO} (data kindly provided by Durouchoux,
	private communication) and so these molecular clouds are
	aligned with the precession cone axis in projection onto the
	sky plane. Their CO(1-0) line Doppler velocity is $\sim
	50$\,km.s$^{-1}$ which corresponds to the same distance of
	$\sim 3.5$\,kpc from the Earth as W50. The limits of these
	clouds are not known because of the limited field observed, so
	we do not know if these two clouds are linked or not. Somme
	data still have to be reduced and may answer to this
	question. If not, we will carry a new observation campaign.
\vspace*{-0.4cm}\\
	   \indent Figure~\ref{figIRCOzoom} zooms on these two CO emitting
	regions. Their shapes are similar to the ISOCAM shapes, so the
	IR and millimetre emissions are unlikely coincident by chance
	but must be physically linked. So the observed 15\,$\mu$m
	emissions are due to regions lying inside W50 and including
	molecular clouds. Knot 2 has also been observed for the
	CO(2-1) line which emiting region has the same shape as the
	CO(1-0) one. This may indicate a shock excited region, the
	study of which is in progress. Thus the nature of the IR emission is
	still uncertain. It could be simply thermal emission from
	young stars regions still embedded in their dust clouds. More
	interesting it could be thermal emission from dust heated by
	the jet interaction with molecular clouds, the high energy
	particles of the jet hitting the CO and dust grains. Otherwise,
	the mid-infrared emission could be of synchrotron nature,
	coming from the very energetic particles of the jet,
	reaccelerated by shocks with the western lobe denser
	medium. These particles would lose their energy by synchrotron
	radiation in X-ray, then in IR as they travel further from
	SS\,433.

\vspace*{-0.0cm}\section{Analogy with extragalactic cases}
	\indent The case of SS\,433 and W50 is interesting in the
	context of the quasar\,/\,microquasar global analogy. Two
	examples can illustrate this, showing as for SS\,433\,/\,W50
	knotty structures in their jet, with X-ray and radio emission
	not spatially coincident: the quasar 3C273 (see Marshall {\em et
	al.}\cite{Marshall} 2001) and the FR\,II radio galaxy Pictor A
	(see Wilson {\em et al.}\cite{Wilson} 2001).

\section{Conclusion}
	\indent I have shown indications of SS\,433 jet interaction
	with the medium of W50 western lobe: the radio lobes lie
	within the relativistic jets precession cone, the observed IR
	hotspots and extended emission are coincident with molecular
	clouds and aligned with the jet axis, and there is a possible
	link with the X-ray emission. The mid-IR emission could be
	either thermal or not (see discussion in
	section~\ref{sectIRmm}).  The prospects to answer this
	question are new observations: with XMM-Newton and Chandra to
	determine the X-ray emission nature (thermal or not, or both),
	in near-IR to find possible embedded stars, in millimeter to
	complete the CO cloud mapping and find traces of shocks, and
	finally with INTEGRAL, SS\,433 being part of the guaranted
	time, a detection in soft $\gamma$-ray would prove that knots
	are sites for particle acceleration. Any results about
	SS\,433\,/\,W50 should be regarded in the context of the
	quasar\,/\,microquasar analogy.

\section*{Acknowledgments}
	\vspace*{-0.2cm}The ISOCAM data presented in this paper was
	analysed using ``CIA", a joint development by the ESA
	Astrophysics Division and the ISOCAM Consortium led by the
	ISOCAM PI, C. Cesarsky, Direction des Sciences de la Matiere,
	C.E.A., France.

\end{document}